\documentclass[aps,prd,preprint]{revtex4-1}
\usepackage{graphicx}
\usepackage{bm}

\newcommand{\Y}{\Y_{lm}}

\newcommand{\be}{\begin{equation}}
\newcommand{\ee}{\end{equation}}
\newcommand{\Be}{\begin{eqnarray}}
\newcommand{\Ee}{\end{eqnarray}}

\newcommand{\f}{\frac}
\begin{document}
\pagestyle{plain}

\title{The Cosmological Model with a Wormhole and Hawking Temperature near Apparent Horizon}

\author{Sung-Won Kim}
\email[email:]{sungwon@ewha.ac.kr}
\affiliation{Department of Science Education, Ewha Womans University, Seoul 03760, Korea}

\begin{abstract}
In this paper, a cosmological model with an isotropic form of the Morris-Thorne type wormhole was derived
in a similar way to the McVittie solution to the black hole in the expanding universe.
By solving Einstein's field equation with plausible matter distribution, we found the exact solution of the wormhole embedded in Friedmann-Lema\^{\i}tre-Robertson-Walker universe.
We also found the apparent cosmological horizons from the redefined metric and analyzed the geometric natures, including causal and dynamic structures.
The Hawking temperature for thermal radiation was obtained by the WKB approximation using the Hamilton-Jacobi equation and Hamilton's equation, near the apparent cosmological horizon.
\end{abstract}
\maketitle

\section{introduction}

The solution to the black hole embedded in expanding universe has been familiar to relativists and
 cosmologists for a long time since McVittie derived a model \cite{MV}.
The cosmological black hole solutions are more like a realistic model.
       Recently, the solutions also has attracted us because of the role of black hole in the expanding cosmological model. The evidences for the accelerating universe and dark energy forced the study of the cosmological black hole model.
The interaction of black holes with dark energy distributed over the universe can be one of the most important issues. Moreover, they can show a generalized theory of global and local physics, that is interested in the unification of interactions \cite{FJ}.

After McVittie solution there were several models of black hole in the universe. McVittie spacetime \cite{MV} was a spherically symmetric, shear free, perfect solution and was asymptotically Friedmann-Lema\^{\i}tre-Robertson-Walker (FLRW) model. In this model, matters were isotropically distributed and there was no-accretion onto the black hole centered at the FLRW universe.
As generalizations of the McVittie solution, there were solutions of the charged black hole in expanding universe \cite{FZP,GZ}. Faraoni and Jacques obtained a generalized McVittie solution without no-accretion condition \cite{FJ}.
  Sultana-Dyer \cite{SD} got the extension of the geometry generated by conformally transforming the Schwarzschild metric with the scale factor of flat FLRW universe. Kottler \cite{Ko} derived the solution of the Schwarzschild black hole in de Sitter background.

The research on wormhole is also an important issue in study of spacetime physics.
The wormhole usually consists of exotic matter which satisfies the flare-out condition and violates weak energy condition \cite{MT,MTY}, even though there have been attempts to construct wormhole with non-exotic matter \cite{example}.
There were also solutions of cosmological wormhole model as well as the cosmological black hole solutions. There was the solution of a wormhole in inflationary expanding universe model \cite{roman}. In this solution, the wormhole throat inflates at the same rate as that of the scale factor. Also there was a wormhole solution in FLRW cosmological model \cite{kim}.
The solution also showed the expansion of the wormhole throat at the same rate as that of the scale factor.
 Hochberg and Kepart tried to extend the  Visser type wormhole into a surgical connection of two FLRW cosmological models \cite{hochberg}.
Similarly there was a solution for the connection of two copies of Schwarzschild-de Sitter type wormhole as the cosmological wormhole model \cite{kim3}.  There was a research on quantum cosmological approach by considering wave function of the de Sitter cosmological model with a wormhole \cite{Barcelo}.
Recently there was a cosmological wormhole solution \cite{MED} as a generalization of MT wormhole in FLRW universe, but there was a weak point that Einstein's equation could not be guaranteed.

First of all, it is necessary to find the exact solution of the cosmological wormhole model satisfying Einstein field equation. For reasons similar to black holes, the exact wormhole model embedded in the universe is very interesting to us. They will provide a lot of information to understand the  relationship
between wormhole matter and the background spacetime.
Most of the previous cosmological wormhole solutions originated from the spacetime assumed to be plausible models.
If a wormhole were in the expanding universe, it would interact with dark energy in some way that would be considered. The impacts on the spacetime are also very interesting. So we need to find the exact solution of wormhole universe to see its influence to the evolution of spacetime.

In the spacetime structure with strong gravity, the Hawking temperature is one of the issues focused on gravitation problem dealing with quantum phenomena.
The Hawking temperature, derived from the definition of surface gravity at event horizon, is a good example of the semi-classical handling of quantum gravity. Usually the temperature is calculated from the surface gravity defined by Killing vector in static case.
In the dynamic case, such as a black hole or a wormhole in expanding universe, the surface gravity at the event horizon is not constant.
In this case, we need to adopt the Kodama vector \cite{kodama} instead of the Killing vector, and try to find the corresponding physical quantities for the spacetime \cite{hayward}.
By using the WKB approximation to the tunneling method, the Hawking temperature is derived by comparing the thermal distribution and probability amplitude from Hamolton-Jacobi eqution \cite{hayward2}. There is also another way to find the probability amplitude from Hamilton's equation, which was designed by Parikh and Wilczek \cite{PW}. There is an example of Hawking temperature at apparent horizon of the FLRW model in both methods \cite{CCH09}.

In this paper, we have found the exact solution of the wormhole embedded in FLRW universe, the locations and the existence conditions of apparent horizon. The influence of wormhole matter to the structure of the apparent horizons was studied. The Hawking radiation was also discussed, and the temperature was also derived near the apparent horizon.

\section{cosmological wormhole model}

\subsection{Isotropic Wormhole}

Now we derive the exact model of the wormhole embedded in FLRW cosmology.
First, we need to find the isotropic form of the wormhole model in order to derive the wormhole solution embedded in a cosmological model because of the isotropy of the cosmological models in this paper.
The Morris-Thorne type wormhole (MT-wormhole) is given by \cite{MT}
\be
ds^2 = - e^{2\Phi} dt^2 +   \f{1}{1-b(r)/r} dr^2 + r^2 d\Omega^2,   \label{mt_worm}
\ee
where $\Phi(r)$ is red-shift function and $b(r)$  is the shape function. The geometric unit, that is, $G=c=\hbar=1$ is used here. The radial coordinate $r$ is in the range of $b<r<\infty$. Two functions $\Phi(r)$ and $b(r)$ are restricted by the `flare-out condition' to maintain the shape of the wormhole.
Because the wormhole has the structure that prevents the existence of the event horizon,
wormhole can be used for two-way travel.
Since MT-wormhole is spherically symmetric form, we introduce the new coordinates $(\tilde{t}, \tilde{r})$ to define the isotropic form of a wormhole as
\be
ds^2 = - A^2 d\tilde{t}^2 + B^2 ( d\tilde{r}^2 + \tilde{r}^2 d\Omega^2 ). \label{iso}
\ee
In this paper, we treat the example of $b(r) =  \f{b_0^2}{r}$ and $e^{2\Phi}=1$
to see the nature of wormhole geometry more simply. Then $B$ becomes
\be
B = \f{2}{1+\sqrt{ 1 - \f{b_0^2}{r^2} }},
\ee
where the integration constant is determined by the asymptotically flat condition.
The new coordinate $\tilde{r}$ is given in terms of old coordinates $r$ as
\be
\tilde{r} = \f{r}{B} = \f{1}{2}( r+\sqrt{ r^2 - b_0^2 }),~~~~~(b_0/2 < {\tilde{r}} < \infty). \label{r-r0}
\ee
The old coordinate $r$ and the radial factor $B$ are given in terms of $\tilde{r}$ as
\be
r = \tilde{r}+ \f{b_0^2}{4\tilde{r}}, \qquad
B = \f{r}{\tilde{r}} = \left( 1 + \f{b_0^2}{4\tilde{r}^2} \right). \label{transform}
\ee
Thus the isotropic form of the static wormhole is
\be
ds^2 = - d\tilde{t}^2 + \left( 1 + \f{b_0^2}{4\tilde{r}^2} \right)^2 ( d\tilde{r}^2 + \tilde{r}^2 d\Omega^2 ),
\ee
for the ultra-static case.
In this isotropic wormhole model, the metric does not diverge at $\tilde{r}=b_0/2$, while it diverges at  $r=b_0$ in MT-wormhole.
This is because this non-diversity is removed by the coordinate transformation.
We can also transform the matter part of the old model \cite{kim2} into new one by using the relationship between two coordinates (\ref{transform}).
For wormhole solution, the matter components in new coordinates are
\Be
{T^{\tilde{0}}}_{\tilde{0}} &=& - \f{4^4b_0^2{\tilde{r}}^4}{8\pi(b_0^2+4{\tilde{r}}^2)^4} =  \rho_w,
\label{eq:rho}\\
{T^{\tilde{1}}}_{\tilde{1}} &=& - \f{4^4b_0^2{\tilde{r}}^4}{8\pi(b_0^2+4{\tilde{r}}^2)^4} = - \tau_w, \label{eq:p1}\\
{T^{\tilde{2}}}_{\tilde{2}} &=& + \f{4^4b_0^2{\tilde{r}}^4}{8\pi(b_0^2+4{\tilde{r}}^2)^4} =  P_w,\label{eq:p2} \\
{T^{\tilde{3}}}_{\tilde{3}} &=& + \f{4^4b_0^2{\tilde{r}}^4}{8\pi(b_0^2+4{\tilde{r}}^2)^4} =  P_w, \label{eq:p3}
\Ee
where $\rho_w, \tau_w,$ and $P_w$ are wormhole energy density, tension, and pressure, respectively.
The negative density is still required for the isotropic form of wormhole model.

\subsection{Exact Solution}

From now on, we use un-tilde coordinate in isotropic form for convenience if only there are no confusions in the rest of this paper.
The FLRW spacetime in isotropic form is given by \cite{MV}
\be
ds^2 = - dt^2 + \f{a^2(t)}{(1+kr^2)^2} ( dr^2 + r^2 d\Omega^2 ).
\ee
Here $a(t)$ is the scale factor and $k=1/4{\cal R}^2$, where ${\cal R}$ is the curvature, and $k$ goes zero in case of flat FLRW spacetime.  We start from the general isotropic metric element to see the unified wormhole in FLRW cosmological model as
\be
ds^2 = - e^{\zeta(r,t)} dt^2 + e^{\nu(r,t)} ( dr^2 + r^2 d\Omega^2 ).
\ee
Similar to the McVittie solution, the following matter distribution in the universe is assumed:
A spherical symmetric distribution of matter around the origin where there is a wormhole, no flow of the matter as a whole either towards or away from the origin, and isotropic pressure of matter in the universe.
In addition to these assumptions, we add one more assumption that the local and the global matters are separate. Since the wormhole is a localized object, the matter of wormhole is separated  from the cosmic matter, which is distributed over the universe.
When we adopt the ansatz, the cosmological matter term is time-dependent and isotropic, while the wormhole matter term depends only on space and not necessarily isotropic as,
\Be
a^2(t)\rho(r,t) &=& a^2(t)\rho_c(t) + \rho_w(r), \\
a^2(t)p_1(r,t) &=& a^2(t)p_{1c}(t) + p_{1w}(r), \label{ans_p1}\\
a^2(t)p_2(r,t) &=& a^2(t)p_{2c}(t) + p_{2w}(r), \label{ans_p2}\\
a^2(t)p_3(r,t) &=& a^2(t)p_{3c}(t) + p_{3w}(r),
\Ee
which were already used in the previous wormhole cosmological model \cite{kim}. If we estimate the orders of magnitude, $\rho_cc^2 \sim 9 \times 10^{-10}{\rm J/m^3} $ over the universe and $\rho_wc^2 \sim 6 \times 10^6 (\f{10 {\rm m}}{r})^4(\f{b_0}{10 {\rm m}})^2 {\rm J/m^3}$ near throat, where $\rho_c$ is roughly the critical density for flat universe.
The wormhole matter is concentrated at or near throat and it decreases rapidly as $1/r^4$, as wee see in (\ref{eq:rho})-(\ref{eq:p3}).
Thus $\rho_c \ll \rho_w$ near throat at least.

Einstein's equation is given by
\[
G_{\alpha\beta} = \kappa T_{\alpha\beta},
\]
where $\kappa = 8\pi$.
The non-zero components of Einstein tensor ${G^\mu}_\nu$ are
\Be
{G^0}_0 &=&  \f{1}{4r} \{ [ (8\nu' + 4\nu'' r + \nu'^2 r ) e^{-\nu+\zeta} - 3 \dot{\nu}^2 r ] e^{-\zeta} \}, \\
{G^0}_1 &=& \f{1}{2}(2\dot{\nu}' - \dot{\nu}\zeta') e^{-\zeta},\\
{G^1}_1 &=&  \f{1}{2r}\{ [r(-2\ddot{\nu} + ( -\f{3}{2}\dot{\nu} + \dot{\zeta}) \dot{\nu} ) e^{-\zeta+\nu} + 2\nu' + 2\zeta' + \zeta'\nu'r + \f{1}{2}r \nu'^2 ] e^{-\nu} \},\\
{G^1}_0 &=& \f{1}{2}(-2\dot{\nu}' + \dot{\nu}\zeta') e^{-\nu},\\
{G^2}_2 &=& {G^3}_3 = \f{1}{4r} \{ [  2\zeta' +2\nu' + 2\nu''r + 2\zeta''r + \zeta'^2 r] e^{-\nu}
+2r( -2\ddot{\nu} -\f{3}{2}\dot{\nu}^2 + \dot{\zeta}\dot{\nu} ) e^{-\zeta+\nu} \}.
\Ee
Here dot denotes the derivation with respect to $t$ and prime denotes the derivative with respect to $r$.

For the case of ultra-static observer ($\zeta = 0$), ${G^1}_0=0$ becomes
\be
\dot{\nu}' = 0. \label{eq-nu}
\ee
The general solution to (\ref{eq-nu}) is
\be
\nu(r,t) = \alpha(t) + \beta(r) ~~~\mbox{or}~~~ e^{\nu(t,r)} = e^{\alpha(t)} e^{\beta(r)} .\label{nu}
\ee
The time-dependent part $\alpha(t)$ relates with the scale factor, $e^{\alpha(t)}\equiv a^2(t)$,
while $\beta(r)$ determines curvature of background and wormhole spacetime, with the boundary condition as
\be
 \left\{ \begin{array}{lcll}
                 e^{\beta(r)} = (1 + \f{b_0^2}{4r^2})^2 & \mbox{or} &\beta(r) = 2\ln(1 + \f{b_0^2}{4r^2}),
                  &~~\mbox{for}~r \rightarrow b_0/2, \\
                 e^{\beta(r)} = (1+ kr^2 )^{-2} & \mbox{or} &\beta(r) = -2\ln(1+ kr^2 ),
                  & ~~\mbox{for}~r \rightarrow \infty.
                 \end{array}
                 \right.
\label{bc}
\ee
Since these two boundary conditions are dominant in the local and global regions, we can find the unified form of $e^{\beta(r)}$.
The limit $r \rightarrow b_0/2$ (or $k=0$) represents the value of the wormhole throat, which is a local solution to the wormhole. The limit $r \rightarrow \infty$ (or $b_0=0$) is an asymptotic solution away from the origin, that is, the metric of the FLRW cosmological model.

When we compare ${G^1}_1$ and ${G^2}_2$, we get
\be
[\nu'' - \f{1}{r}\nu' - \f{1}{2}(\nu')^2]e^{-\nu} = 2\kappa ( p_2  - p_1 ). \label{g11g22}
\ee
For the cosmological term, $p_{1c} = p_{2c}$ due to the isotropy, while $p_{1w} \neq p_{2w}$ for the wormhole case.
In case of the wormhole, $p_{1w}$ is the negative value of the tension.
If we use ansatz to separate temporal parts, as in the previous case (\ref{ans_p1}) and (\ref{ans_p2}), the equation contains only pure spatial parts as
\be
[\beta'' - \f{1}{r}\beta' - \f{1}{2}(\beta')^2]e^{-\beta} =
2\kappa(p_{2w}-p_{1w})=\f{4b_0^2}{r^4(1+b_0^2/4r^2)^4}.\label{inhom}
\ee
Here we use the wormhole matter components of (\ref{eq:p1}) and (\ref{eq:p2}).
The general solution to this inhomogeneous differential equation is the solution to the homogeneous
equation plus the special solution.
For the black hole cosmological model, the right hand side is canceled and the equation is homogeneous \cite{MV}. The solution for this wormhole is $\beta = \beta_c + \beta_w $, where $\beta_c$ is the same as that of the black hole cosmology \cite{MV} and $\beta_w$ is the special solution that satisfies (\ref{inhom}) and boundary condition (\ref{bc}). The physical meaning of $\beta_c$ is the global cosmological background which dominates at far region. The meaning of $\beta_w$ is the local wormhole
solution which dominates near origin. The global cosmological solution is
\be
\beta_c = -2 \ln ( kr^2 + 1)
\ee
with the boundary condition for FLRW model (\ref{bc}). The local wormhole solution $\beta_w$ (special solution) that satisfies (\ref{inhom}) with the boundary condition  (\ref{bc}) becomes
\be
\beta_w = 2 \ln \left( 1 + \f{b_0^2}{4r^2} \right).
\ee
Thus the general solution to the equation (\ref{nu}) is
\be
e^{\nu(r,t)} = e^{\alpha(t)+\beta_c(r)+\beta_w(r)}= \f{e^{\alpha(t)}}{( kr^2 + 1)^2} \left( 1 + \f{b_0^2}{4r^2} \right)^2 \label{g_rr}
\ee
and the cosmological wormhole solution finally becomes
\be
ds^2 = - dt^2 + \f{a^2(t)}{( kr^2 + 1)^2} \left( 1 + \f{b_0^2}{4r^2} \right)^2 (dr^2 + r^2 d\Omega^2). \label{worm-cos}
\ee
When we inverse-transform this spacetime, at least in $k=0$ case,
 into the spherically symmetric type, it returns to the MT-wormhole times the time-dependent scale factor $a^2(t)$.
For flat FLRW universe, the spacetime is same form as the previous one \cite{kim}.
\be
a^2(t)\left[ \f{dr^2}{1-\f{b(r)}{r}} + r^2 d\Omega^2 \right].
\ee
Now we found the exact solution of the wormhole in FLRW cosmological model.
In this model, matters are separated into time-dependent part for cosmological background $\rho_c(t)$, and space-dependent part for wormhole $\rho_w(r)$.
Apparently  the cosmic part is coupled with wormhole part in spacetime metric of the cosmological wormhole as in (\ref{g_rr}). However, the wormhole matter does not affect cosmological background, since $\rho_w$ does not affect the term $(1+kr^2)$ determined by  $\rho_c(t)$.

\section{Apparent Horizons and Causal Structures}

To understand the causal structure of the spacetime, we should find and analyze the apparent horizons.
If we redefine the new coordinate $R$ from the spacetime (\ref{worm-cos}) as
\be
R \equiv a \left( \f{1+b_0^2/4r^2}{1+kr^2} \right) r  = a(t)A(r),
\ee
the metric can be rewritten as $ds^2 = h_{ab}dx^adx^b + R^2 d\Omega^2$, where $x^a=(t,R), h_{ab} = {\rm diag} (-1, a^2(1+b_0^2/4r^2)/(1+kr^2))$.
The location of the apparent horizon is defined from the relation $h^{ab}\partial_aR\partial_bR=0$.
When we see the metric, the apparent horizon is located at the position which satisfies the relation as
\be
\Delta \equiv 1 - \f{H^2R^2}{r(R)^2J(R)^2} = 0, \label{hor1}
\ee
where
\[
\f{\dot{a}}{a}=H,~~~\f{A'}{A}=J=\f{1}{r} - \f{b_0^2/2r^3}{1+b_0^2/4r^2}-\f{2kr}{1+kr^2}.
\]
In the redefined coordinates system, $\Delta$ appears as the coefficient of $g_{tt}$ and $g_{RR}$
for Schwarzschild-like metric.
Simply we here see the $k=0$ case, which is the wormhole spacetime in flat FLRW universe. In this case,
$r$ is given in terms of $R$ by
\be
r  = \f{R}{2a} \pm \sqrt{\left( \f{R}{2a} \right)^2 - \left( \f{b_0^2}{4} \right)}.\label{r_to_R}
\ee
There are two relations between $r$ and $R$, but the plus sign is chosen only due to the positive nature of $r$. When $R=a(t)b_0$, two relations meet and $r=R/2a$.
By putting this (\ref{r_to_R}) into relation (\ref{hor1}), we get the location of
the apparent horizons at
\be
R_\pm = \f{1}{\sqrt{2H^2}} [  1 \pm \sqrt{1-(2b(t)H)^2}]^{1/2} < \f{1}{H} \equiv R_H, \label{horizon}
\ee
where $b(t)\equiv a(t)b_0$ is the size of the wormhole throat scaled by $a(t)$ and is equal to the minimum of $R$ at $r=b_0/2$.  $R_H=1/H$ is the apparent cosmological horizon of FLRW model in absence of wormhole ($b_0=0$) and is called `Hubble horizon'. When $b_0=0$, the apparent horizons approach $R_+=R_H$ and $R_-=0$, respectively.

\begin{figure}[h]
\includegraphics[height=6.5cm]{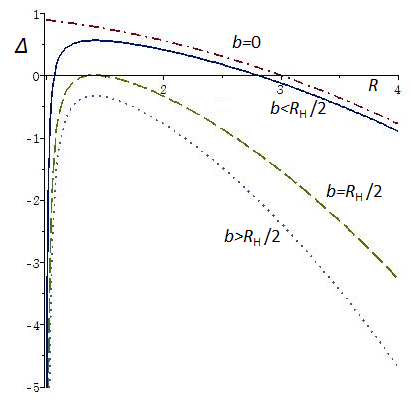}
\caption{The function $\Delta=(1-R^2H^2/r^2J^2)=(R^2-b^2-R^4H^2)/(R^2-b^2)$. Here we set $b=1$.
The three cases for the number of solutions to $\Delta=0$ are shown. Two solutions are shown for $b<R_H/2$ (solid line), one solution is shown for $b=R_H/2$ (dashed line), and no solution for $b>R_H/2$ (dotted line). The dash-dot line is the case of $b=0$.}
\end{figure}

As shown in Fig. 1, the equation $\Delta=0$ provides two, one, or zero solutions, depending on the relative values of $b(t)$ for $1/(2H)=R_H/2$.  Since we treat the isotropic coordinate in cosmological wormhole, the minimum is the half value of the spherically symmetric case as in (\ref{r-r0}) and the comparison of $R_{\rm min}$ to $R_H/2$ is reasonable.
We will consider the three cases for the solution to $\Delta=0$.

Case 1: $b(t)<R_H/2$. In this case there are two horizons: the larger is $R_+$ and the smaller is $R_-$. The former is the cosmological horizon and the latter is the size of the wormhole throat in the expanding universe. The fact that $R_+ < R_H$ means that the wormhole reduces the Hubble horizon to a smaller size and $b(t) < R_-$ indicates that the cosmological background enlarges the wormhole throat size.
Only in $R_-<R<R_+$, the coordinate $t$ is timelike and $R$ is spacelike. In the region $b<R<R_-$ or $R_+<R<R_H$, the role of time and space coordinates is reversed. The coordinate $R$ is limited to $R>R_-$ similar to a static wormhole, so we will do not consider the region smaller than $R_-$.

Case 2: When $b(t)=R_H/2$, two horizons coincide and there is only one horizon, such as $R_\pm=1/(\sqrt{2}H)$.
In this case the metric coefficient $\Delta$ is negative except for that point.

Case 3: If $b(t) > R_H/2$, there is no apparent horizon. In this case, the coefficient $\Delta$ is always negative (never zero). Thus the coordinate $t$ is spacelike and $R$ is timelike. There is no regular region.
This relationship shows that the size of the scaled wormhole throat can't be larger than the half of the FLRW Hubble horizon, if horizons should exist.

Fig. 2 shows the apparent horizons in the case of $b(t) \le R_H/2$ at a given time as a function of $b(t)$.
Three horizons, Hubble horizon without wormhole ($R_H$) and two horizons with wormhole
($R_\pm$) at a constant time $t_1$, are depicted in the figure. As mentioned above, when $b(t_1) = R_H/2$, two horizons meet.
At the limit of $b \rightarrow 0$, $R_+$ approaches the FLRW horizon $R_H$ and $R_-$ goes to zero. At the limit of $H \rightarrow 0$, $R_-$ approaches $b(t_1)$.

\begin{figure}[h]
\includegraphics[height=6.5cm]{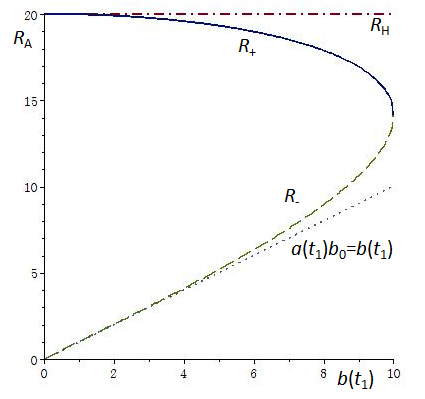}
\caption{The apparent horizons ($R_\pm$) and the Hubble horizon ($R_H$) at a constant time $t_1$ in terns of $b$ are shown. Solid line is $R_+$, dashed line is $R_-$, dash-dot line is $R_H$ which is asymptote of $R_+$, and the dotted line is the asymptote of $R_-$.
Here we set $H=0.05$ at a constant time.}
\end{figure}

The apparent horizons $R_\pm$ are produced from the wormhole matter $\rho_w$ of $b_0$ and its coupling to the cosmic matter $\rho_c$ of $a(t)$. The equation (\ref{eq:rho}) shows that the $b_0$ can be expressed by the peak of wormhole energy density. This value is equal to the energy density at throat of the wormhole as
\be
b_0^2 = \f{1}{8\pi}\left.\f{1}{|\rho_w(r)|}\right|_{r=b_0/2} \equiv \f{1}{8\pi}\f{1}{|\rho_w^p|}, \label{b_rho}
\ee
where $\rho_w^p$ is the peak of $\rho_w(r)$. The most part of energy is distributed near the throat of the wormhole. The term $a^2H^2$ is determined by the distribution of the cosmic matter $\rho_c$, for example,
\be
a^2H^2 = \f{8\pi}{3}\rho_{c0}^{\f{2}{3(1+\omega)}}\rho_c^{\f{1+3\omega}{3(1+\omega)}}
\ee
in case of $P_c=\omega\rho_c$, where $\rho_{c0}$ is the current mass density of the cosmic matter.
Thus the term in square root of (\ref{horizon}) is
\be
1-4b_0^2a^2H^2 
\simeq  1- \f{4}{3}\left(\f{\rho_{c0}}{|\rho_w^p|} \right) \geq 0 \label{root}
\ee
in case of $\rho_c\approx\rho_{c0}$. The apparent horizons are determined by the peak value of the energy density of the wormhole. When $|\rho_w^p|$ is large, $b_0$ is smaller and the horizons are separated. The larger the peak value, the closer the horizon $R_+$ approaches $R_H$ and $R_-$ approaches zero. As $|\rho_w^p|$ becomes smaller, $b_0$ becomes larger. Because of the positivity of (\ref{root}), the horizons disappear when $|\rho_w^p|=\f{4}{3}\rho_{c0}$ (See Fig.~3). 
The exotic matter shrinks the horizons of the universe to a smaller size in both microscopic and macroscopic directions.
In Fig. 2 and 3, the horizons appear and disappear in reverse, because $b_0$ and $|\rho_w^p|$ have a relationship (\ref{b_rho}).

\begin{figure}[h]
\includegraphics[height=6.5cm]{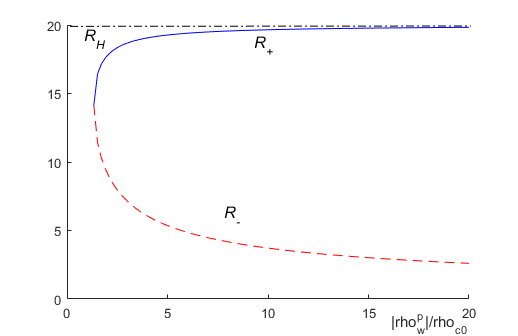}
\caption{The apparent horizons ($R_\pm$) and the Hubble horizon ($R_H$) in terms of $|\rho_w^p|/\rho_{c0}$ are shown. Solid line is $R_+$, dashed line is $R_-$, and dash-dot line is $R_H$.
The apparent horizons begin to appear at  $|\rho_w^p|/\rho_{c0}=4/3$.
Here we set $H=0.05$.}
\end{figure}

In Schwarzschild-like coordinate, the Misner-Sharp-Hermandez mass is derived as usual,
\be
M_{\rm MSH} = H^2 \f{R^3}{2(1-b^2/r^2)} = \f{4\pi}{3}R^3 \rho_c [1-(b/R)^2]^{-1}
\ee
from the definition
\[
1- \f{2M_{\rm MSH}}{R} = 1 - \f{H^2R^2}{r^2J^2}.
\]
The value calculated here is modified by a factor of $[1-(b/R)^2]^{-1}$ compared to that of the FLRW cosmological model without any wormhole, the value of which is the mass of sphere of radius $R$ filled with cosmic matter $\rho_c$.

\section{Hawking Temperature}

We can see the quantum nature of the spacetime by calculating the Hawking temperature near horizon.
In order to discuss the Hawking radiation, we use coordinates $(t,R)$ to rewrite the metric as
\be
ds^2 = - \left( 1 - \f{R^2/(R_+^2+R_-^2)}{1-b^2/R^2} \right) dt^2
-\f{2HR}{1-b^2/R^2} dtdR + \f{1}{1-b^2/R^2} dR^2 + R^2 d\Omega^2,
\ee
where $b=a(t)b_0$. The Kodama vector is
\be
K^a \equiv - \varepsilon^{ab}\partial_bR =  \sqrt{1-\f{b^2}{R^2}} \left( \f{\partial}{\partial t} \right)^a,
\ee
where $\varepsilon^{ab}= \f{1}{ \sqrt{1-{b^2}/{R^2}}}(dt)_a \wedge (dR)_b$.
We should consider a particle of mass $m$ moving radially under the background of FLRW wormhole spacetime.
The Hamilton-Jacobi equation is
\be
g^{\mu\nu}\partial_\mu{\bf S}\partial_\nu{\bf S} + m^2  = 0.
\ee
One can define the energy $\omega$ and momentum $k_R$ with the Kodama vector similar to the case of FLRW univere \cite{CCH09} as
\be
\omega  = -K^a\partial_a{\bf S} = - \sqrt{1-\f{b^2}{R^2}}\partial_t{\bf S},
\qquad
k_R = \left( \f{\partial}{\partial R} \right)^a\partial_a{\bf S} = \partial_R{\bf S}.
\ee
Therefore, the action ${\bf S}$ can be written as
\be
{\bf S} = - \int \f{\omega}{\sqrt{1-b^2/R^2}} dt + \int k_R dR.
\ee
The Hamilton-Jacobi equation with the action is
\be
-\f{\omega^2}{1-b^2/R^2}+
\f{2HR\omega}{\sqrt{1-b^2/R^2}}k_R +
\f{(R_+^2-R^2)(R^2-R_-^2)}{(R_+^2+R_-^2)R^2}k_R^2
+ m^2 =0,
\ee
and the solution to $k_R$ is
\be
k_R = \f{-HR \pm\sqrt{H^2R^2+\lambda[1-\f{m^2}{\omega^2}(1-b^2/R^2)]}}
{\sqrt{1-b^2/R^2}(R_+^2-R^2)(R^2-R_-^2)}R^2(R_+^2+R_-^2)\omega,
\ee
where $\lambda=\f{(R_+^2-R^2)(R^2-R_-^2)}{(R_+^2+R_-^2)R^2}$.
We choose minus sign for incoming mode since the observer is inside the apparent horizon
similarly to the FLRW case \cite{CCH09}.
Through the contour integral, we get
\Be
{\rm Im} ~{\bf S} &=& {\rm Im} ~\int \f{-HR -\sqrt{H^2R^2+\lambda[1-\f{m^2}{\omega^2}(1-b^2/R^2)]}}
{\sqrt{1-b^2/R^2}(R_+^2-R^2)(R^2-R_-^2)}R^2(R_+^2+R_-^2)\omega \nonumber \\
&=&  \pi R_+ \omega.
\Ee
Here we integrate out over the region larger than $R_-$ such that there is only one pole on the contour.
In the WKB approximation, the emission rate is proportional to the square of the tunneling amplitude, $\Gamma \propto \exp(-2~ {\rm Im} ~{\bf S} )$.
By comparing with the form of the thermal spectrum $\Gamma \sim \exp(-\omega/T)$,
we thus obtain the Hawking temperature as
\be
T= \f{1}{2\pi R_+}. \label{temp}
\ee

We can also derive the Hawking temperature using the Hamilton's equation alonng the way by Parikh and Wilczek \cite{PW}.
As the case of de Sitter space \cite{Parikh} and FRW spacetime \cite{CCH09}, we will take the $s$-wave approximation of a massless particles tunneling across the horizon. The emission rate can be related to the imaginary part of the action of a system.
The radial null geodesic of the metric is
\be
\dot{R} = HR \pm \sqrt{(HR)^2+(1-b^2/R^2 - R^2/(R_+^2+R_-^2))},
\ee
where we take the minus (-) sign which corresponds to a incoming null geodesic.
The imaginary part can be
\be
{\rm Im} ~{\bf S} = {\rm Im}\int_{R_i}^{R_f} p_R dR = {\rm Im}\int_{R_i}^{R_f} \int_0^{p_R}
dp_R' dR,
\ee
where $p_R$ is the radial momentum, $R_i$ is the initial position, slightly outside the apparent horizon and $R_f$ is a classical  turning point. By the Hamiltonian equation,
\[
\dot{R} = \f{\partial \hat{H}}{\partial p_R} = \f{d\hat{H}}{dp_R} \left.\right|_R.
\]
Here $\hat{H}$ is the Hamiltonian of the particle, the generator of the cosmic time $t$ as we see in the action.
We can calculate the imaginary part of the action as
\Be
{\rm Im} ~{\bf S} &=& {\rm Im}\int_{R_i}^{R_f} dR \int d\hat{H} \f{1}{\dot{R}} \nonumber \\
&=& -\omega~ {\rm Im} \int_{R_i}^{R_f}\f{dR}{\sqrt{1-b^2/R^2}(\sqrt{(HR)^2+(1-b^2/R^2-R^2/(R_+^2+R_-^2))} - HR)}\nonumber \\
&=& \pi R_+\omega.
\Ee
From this value, we also get the Hawking temperature as
\be
T = \f{\omega}{2~{\rm Im} ~{\bf S}} = \f{1}{2\pi R_+}.
\ee
The result gives the same Hawking temperature as the previous derived one (\ref{temp}) by Hamilton-Jacobi equation.

\section{conclusion}

In conclusion, we would like to remark the following three points.

First, we derived the exact solution of a wormhole embedded in expanding universe in this paper.
The background universe model was FLRW. There were several solutions based on the generalization of the isotropic wormhole solution. For example \cite{MED}, they have generalized the static wormhole metric as a time-dependent cosmological wormhole metric by introducing two functions $w(t,r)$ and $q(t,r)$ similar to the solution of the charged black hole in the unverse \cite{GZ}, as
\[
\left( 1 + \f{b_0^2}{4r^2} \right) ~~\rightarrow~~
\left( w(t,r) + \f{q(t)}{4r^2} \right)~~\rightarrow~~a(t)\left( \f{1}{1+kr^2}+\f{b_0^2}{4r^2}\right).
\]
Here, $w(t,r)$ is related to the background cosmological model and $q(t)$ is related to the time-dependent wormhole shape.
In this generalization, there is not any coupling term of wormhole and cosmological background.
Moreover, the Einstein's equation was not guaranteed. They simply showed additive generalization in their final solution with the wormhole part and the spatial curvature part of cosmological model.

 Since this was not unique one, we had alternative generalization here as
\[
\left( 1 + \f{b_0^2}{4r^2} \right) ~~\rightarrow~~
w(t,r)\left( 1 + \f{q(t)}{4r^2} \right)~~\rightarrow~~a(t)\left(\f{1}{1+kr^2}\right)\left( 1+\f{b_0^2}{4r^2}\right),
\]
and we got the solutions satisfying Einstein's field equation.
In this case, the multiplicative generalization was shown by the wormhole part and the curvature part of the cosmological model.
There was the coupling term of $k\cdot b_0$, that is spacetime curvature-wormhole coupling term that did not appear in the previous generalization. The coupling refers to local and global physics,
such as unified field theory, which links two interactions of extreme regions.

Second, we also found that the number of apparent horizons of the model are depends on the value of $b(t)H$. The values also shown the physically reasonable regions such as timelike $t$- and spacelike $R$-coordinates. The apparent horizons were similar to the Vaidya-de Sitter spacetime with a cosmological horizon and a black hole horizon \cite{kim4}.
Of course, in the black hole spacetime, the causal structure inside the event horizon was reversed compared to the outer spacetime. However, we did not need the region with smaller radius than the wormhole horizon. As a result, the larger cosmological horizon was reduced by wormhole and the smaller wormhole throat size was enlarged by the cosmological background. 

Third, the Hawking temperature was calculated along the contour of the action from Hamilton-Jacobi equation. Even if we use Hamilton's equation, the result was the same. Along the contour, we adopted only one pole due to the limitation of $R>R_-$, while there was the effective temperature from both temperatures according to  the previous two horizons in Vaidya-de Sitter spacetime \cite{kim4}.

\begin{acknowledgments}
This work was supported by National Research Foundation of Korea (NRF) funded by the Ministry of Education (2017-R1D1A1B03031081).

\end{acknowledgments}


\end{document}